\documentstyle[12pt]{article}

\begin{document}

\mbox{} \hskip 9cm DF/IST-10.2001

\mbox{} \hskip 9cm October 2001

\vskip 0.5cm

\begin{center}
{\bf Magnetic strings in
anti-de Sitter General Relativity} \\
\vskip 1cm 
\'Oscar J. C. Dias \\
\vskip 0.3cm
{\scriptsize  CENTRA, Departamento de F\'{\i}sica,
      Instituto Superior T\'ecnico,} \\
{\scriptsize  Av. Rovisco Pais 1, 1049-001 Lisboa, Portugal.} \\
{\scriptsize  oscar@fisica.ist.utl.pt} \\
\vskip 0.6cm
             Jos\'e P. S. Lemos   \\
\vskip 0.3cm
{\scriptsize  CENTRA, Departamento de F\'{\i}sica,
      Instituto Superior T\'ecnico,} \\
{\scriptsize  Av. Rovisco Pais 1, 1049-001, Lisboa, Portugal.} \\
{\scriptsize  lemos@kelvin.ist.utl.pt} \\
\end{center}
\bigskip

\begin{abstract}
\noindent

We obtain spacetimes generated by  static and spinning 
magnetic string sources in  Einstein
Relativity with negative cosmological constant ($\Lambda<0$). 
Since the spacetime is asymptotically a cylindrical anti-de 
Sitter spacetime, we will be able to calculate the mass, momentum, 
and electric charge of the solutions.
We find two families of solutions, one with longitudinal 
magnetic field and the other with angular magnetic field.
The source for the longitudinal magnetic field can be 
interpreted as composed by a system of two symmetric and 
superposed electrically charged lines with one of the 
electrically charged lines being at rest and the other 
spinning. The angular magnetic field solution can be similarly 
interpreted as composed by charged lines but now one is at 
rest and the other has a velocity along the axis. This solution 
cannot be extended down to the origin. 
\newline

\end{abstract}
\newpage

\noindent
{\small
\section{\bf Introduction}
}
 \vskip 3mm

\noindent
{\small
\subsection{\bf Purpose:}
}
\vskip 3mm

The purpose of this paper is to present asymptotically 
anti-de Sitter spacetimes generated 
by  static and spinning magnetic sources in Einstein
Relativity with negative cosmological constant ($\Lambda<0$) 
and with topology different from spherical.
These magnetic string sources are obtained in the limit 
that the string internal structure is restricted to an 
infinite line 
source. Since
the spacetime is asymptotically anti-de Sitter we
will be able to calculate the mass, momentum, 
and electric charge of the solutions.
We find two families of solutions, one with longitudinal 
magnetic field  [the only non-vanishing component of the 
vector potential is $A_{\varphi}(r)$] and the other with 
angular magnetic field [$A_z(r) \neq 0$].

\vskip 3mm
\noindent
{\small
\subsection{\bf Black strings:}
}
\vskip 3mm

Static and rotating uncharged solutions of
Einstein Relativity with negative cosmological constant
and with planar symmetry (planar, cylindrical and toroidal 
topology) have been 
found by Lemos \cite{Lemos1}. 
Unlike the zero cosmological constant planar case, these 
solutions include the presence of 
black strings (or cylindrical black holes) in the 
cylindrical model, and  of toroidal black holes and 
black membranes in the toroidal and planar models, 
respectively.
Klemm, Moretti and Vanzo \cite{KMVanz} extracted from 
the general Petrov type-D solution a different rotating 
toroidal metric of the theory. Topological multi-tori
black hole solutions have also been studied 
\cite{M_T_1}.

The extension to include the Maxwell field has been done by
Zanchin and Lemos \cite{Zanchin_Lemos} who found
the static and rotating pure electrically charged black 
holes that are the electric counterparts of the cylindrical, 
toroidal and planar black holes found in
\cite{Lemos1}.
The metric with electric charge and zero angular momentum
was also discussed by Huang and Liang \cite{Huang_L}.

Since the discovery of these black holes with
cylindrical, toroidal and planar topology, many
works have appeared dedicated to the study of their
properties.  
As a test to the cosmic censorship and hoop
conjectures, gravitational collapse of these black 
holes in a background with a negative
cosmological constant has been studied by Lemos 
\cite{Lemos3} and Ghosh \cite{Ghosh}. 
The relationship between the topology of the horizon and 
the topology at infinity was established in 
\cite{Top_1} for cylindrical, toroidal 
and planar black holes.
The thermodynamics of asymptotically anti-de Sitter 
spacetimes and, in particular, of toroidal black holes
has been studied in \cite{Term_1}. 
DeBenedictis \cite{Pol_1} has studied scalar
vacuum polarization effects in cylindrical black 
holes.
The supersymmetry properties of toroidal and cylindrical
black holes in anti-de Sitter spacetimes have been 
obtained in \cite{Lemos6} and their 
quasi-normal modes have been 
studied by Cardoso and Lemos \cite{Vitor}.
A more complete and recent review on this 
subject and on its various connections can be found
in \cite{Lemos_Rev}.

\vskip 3mm
\noindent
{\small
\subsection{\bf Bare (or cosmic) strings:}
}
\vskip 3mm

In this paper we are dealing directly with the issue 
of spacetimes generated by string sources in four 
dimensional Einstein theory that are horinzonless and
have nontrivial external solutions. A short review 
of papers treating  this subject follows.
Levi-Civita \cite{L-C} and Marder \cite{Marder} have  
given static uncharged cylindrically and axially symmetric 
solutions of Einstein gravity with vanishing cosmological constant.
Since then, Vilenkin \cite{Vil_1}, Ford and Vilenkin \cite{Ford},
Hiscock \cite{Hiscock}, Gott \cite{Gott_1},
Harari and Sikivie \cite{Harari_S},
Cohen and  Kaplan  \cite{Cohen_K},
Gregory  \cite{Gregory} and
Banerjee, Banerjee and Sen  \cite{Ban} have found similar 
static solutions in the context of 
cosmic string theory. Cosmic strings are topological structures
 that arise
from the possible phase transitions to which the universe
might have been subjected to and may play an important
role in the formation of primordial structures.
These solutions \cite{L-C}-\cite{Ban} have in common a
 geometrical
property. They are all horizonless and the 
corresponding space has a conical geometry, i.e, it is 
everywhere flat except at the location of the line source. 
The space can be obtained from the flat space by 
cutting out a wedge and identifying its edges. 
The wedge has an opening angle which turns to be 
proportional to the source mass.

Electromagnetic strings, i.e, horizonless cylindrically and 
axially symmetric solutions of Einstein-Maxwell gravity 
have also been obtained. Static electrically 
charged solutions with $\Lambda=0$ were found by 
Mukherji \cite{Muk} while static magnetic solutions with 
$\Lambda=0$ have been constructed 
by Bonnor \cite{Bonnor}, Witten \cite{Witten} and
Melvin \cite{Melvin_0}. Superpositions of these solutions
were considered by Safko \cite{Safko}. 
In the context of electromagnetic cosmic strings, 
Witten \cite{E_Witten} has shown that there are cosmic strings, 
known as superconducting cosmic strings, that behave as 
superconductors and have interesting interactions with 
astrophysical magnetic fields (see also \cite{Peter}). 
These strings can also produce 
large astronomical magnetic fields \cite{Larg_B_1}. 
Moss and Poletti \cite{Moss_P} studied the gravitational 
properties of superconducting cosmic strings while 
non stationary superconducting cosmic strings that reduce, 
in a certain limit, to the solutions found by Witten 
\cite{Witten} were obtained by Gleiser and Tiglio \cite{Gleiser_T}. 
Superconducting cosmic strings have also been studied 
in Brans-Dicke theory by Sen \cite{Sen}, and in dilaton 
gravity by Ferreira, Guimar\~aes and 
Helayel-Neto \cite{Ferreira}. 

\vskip 3mm
\noindent
{\small
\subsection{\bf Corresponding three dimensional 
solutions:}
}
\vskip 3mm

The relation between cylindrically symmetric four 
dimensional solutions and spacetimes generated by 
point sources in three 
dimensions has been noticed in many works.
For example, the cylindrical black hole found
in \cite{Lemos1} reduces (through dimensional reduction) 
to a special case of the black holes of a Einstein-Dilaton theory 
of the Brans-Dicke type \cite{Sa_Lemos_Static}. The 
black holes of a Einstein-Maxwell-Dilaton theory 
of the Brans-Dicke type \cite{OscarLemos} contain as a special 
case the three dimensional counterpart of the electrically charged
black hole found in \cite{Zanchin_Lemos}.
Now, the study of horizonless spacetime solutions generated by 
point sources in three dimensions has begun with
Staruszkiewicz \cite{Star} and 
Deser, Jackiw and t' Hooft \cite{DJH_flat} who have studied
the three dimensional counterparts of the four dimensional
solutions found by Marder \cite{Marder}.
Since then, this issue has been object of many studies. For 
reviews see \cite{3D_Review,OscarLemos_2}. 

\vskip 3mm
\noindent
{\small
\subsection{\bf Plan:}
}
\vskip 3mm

The plan of this article is the following. 
Section 2 deals with the longitudinal magnetic field solution 
($A_{\varphi} \neq 0$) and section 3 treats the  
angular magnetic field solution ($A_z \neq 0$).
In section 2.1 we set 
up the action and the field equations. 
In section 2.2, the static longitudinal  
solution is found and we analyse in detail the 
causal and geodesic structure.
 Angular momentum is added in section 2.3.
In section 2.4, we calculate the mass, angular momentum, and 
electric charge of the longitudinal solutions. 
In section 2.5, we give a physical interpretation for the 
source of the longitudinal magnetic field. 
The angular magnetic solution is discussed in section 3. 
Finally, in section 4 we present the concluding remarks.

\vskip 1cm
\noindent
{\small
\section{\bf Longitudinal magnetic field solution}
}
\vskip 3mm

\vskip 0.3cm
\noindent
{\small
\subsection{\bf Field equations}
}
\vskip 3mm

We are going to work with the Einstein-Hilbert action 
coupled to electromagnetism in four dimensions 
with a negative cosmological term 
\begin{equation}
S=\frac{1}{8\pi} \int d^4x \sqrt{-g} 
   {\bigl [} R - 2 \Lambda - F^{\mu \nu}F_{\mu \nu}
  {\bigr ]},                                   \label{ACCAO}
\end{equation} 
where $g$ is the determinant of the metric, 
$R$ is the curvature scalar, 
$\Lambda$ is the negative cosmological constant and 
$F_{\mu\nu} = \partial_\nu A_\mu - \partial_\mu A_\nu$ is the 
Maxwell tensor, with $A_\mu$ being the vector potential. 
We work with units such that 
$G \equiv 1$ and $c \equiv 1$. 

Varying this action with respect to $g^{\mu \nu}$ and 
$F^{\mu \nu}$ one gets the Einstein and Maxwell equations, 
respectively
\begin{eqnarray}
   & &  G_{\mu \nu} + g_{\mu \nu} \Lambda= 8 \pi T_{\mu \nu},         
                               \label{EQUACAO_MET} \\   
   & &  \nabla_{\nu}F^{\mu \nu}=0 \:,   
                                  \label{EQUACAO_MAX} 
\end{eqnarray}
where $G_{\mu \nu}=R_{\mu \nu} -\frac{1}{2} g_{\mu \nu} R$ 
is the Einstein tensor, $\nabla$ represents the covariant 
derivative and 
$T_{\mu \nu}=\frac{1}{4 \pi}(g^{\gamma \sigma}F_{\mu \gamma} 
F_{\nu \sigma}-\frac{1}{4}g_{\mu \nu}F_{\gamma \sigma}
F^{\gamma \sigma})$ is the Maxwell energy-momentum tensor. 

We want to consider now a spacetime which is both static 
and rotationally symmetric, implying the existence of a 
timelike Killing vector $\partial/\partial t$ and 
a spacelike Killing vector $\partial/\partial\varphi$. 
We will work with the following ansatz for the metric 
\begin{equation}
  ds^2 = -\alpha^2 r^2 dt^2 + e^{-2\nu(r)}dr^2    
   +\frac{e^{2\nu(r)}}{\alpha^2} d\varphi^2 + e^{2\mu(r)}dz^2 \:,
                               \label{MET_SYM}
\end{equation}
where the parameter $\alpha^2$ is, as we shall see, an appropriate 
constant proportional to the cosmological constant $\Lambda$. 
It is introduced in order to have metric components with 
dimensionless units and an asymptotically anti-de Sitter 
spacetime.
The motivation for this curious choice for the metric gauge 
[$g_{tt} \propto- r^2$ and 
$(g_{rr})^{-1} \propto g_{\varphi \varphi}$] 
instead of the usual Schwarchild gauge [$(g_{rr})^{-1} =-g_{tt}$ 
and  $g_{\varphi \varphi}=r^2$] comes from the fact that we are
looking for magnetic solutions. Indeed, let us first remember 
that the Schwarzschild gauge is usually an appropriate choice 
when we are interested on electric solutions. 
Now, we focus on the well known fact that the electric field 
is associated with the time component, $A_t$, of the vector 
potential while the magnetic field is associated with
the angular component $A_{\varphi}$. From
the above facts, one can expect that
a magnetic solution can be written in a metric gauge in which
the components $g_{tt}$ and $g_{\varphi \varphi}$ interchange
their roles relatively to those present in the Schwarzschild 
 gauge used to describe electric solutions.  
This choice will reveal to be a good one to find solutions since 
we will get a system of differential equations that have a 
straightforward exact solution. However, as we will see, it
is not the good coordinate system to interpret the solutions.
 
We also assume that the only non-vanishing components of the 
vector potential are $A_t(r)$ and $A_{\varphi}(r)$, i.e. , 
\begin{equation} 
A=A_tdt+A_{\varphi}d{\varphi}\:.  
\label{Potential} 
\end{equation}
This implies that the non-vanishing components of the symmetric 
Maxwell tensor are $F_{tr}$ and $F_{r \varphi}$. Inserting 
metric (\ref{MET_SYM}) into equation (\ref{EQUACAO_MET}) one 
obtains the following set of equations 
\begin{eqnarray} 
  & &  -\frac{\nu_{,r}}{r}- \frac{\mu_{,r}}{r}(1+r\nu_{,r})
      - \Lambda e^{-2\nu}=-8\pi T_{rr} \:,    
                                      \label{MET_11}  \\
  & &   \mu_{,rr}+ (\mu_{,r})^2 +
      \frac{\nu_{,r}}{r}+ \frac{\mu_{,r}}{r}(1+r\nu_{,r})
      + \Lambda e^{-2\nu}
     =8\pi \alpha^2 e^{-4\nu} T_{\varphi \varphi} \:,   
                                        \label{MET_22}  \\ 
  & &   0=8\pi  T_{t \varphi}=2 e^{2\nu}F_{tr}F_{\varphi r} \:,
                                           \label{MET_02} 
\end{eqnarray}
where ${}_{,r}$ denotes a derivative with respect to $r$.
In addition, inserting  metric (\ref{MET_SYM}) into equation 
(\ref{EQUACAO_MAX}) yields 
\begin{eqnarray}
  \partial_r {\bigl [} 
      r e^{\mu}(F^{t r}+F^{\varphi r}) {\bigr ]}=0\:.  
                                 \label{MAX_0} 
\end{eqnarray}

\vskip 0.3cm

\vskip 3cm
\noindent
{\small
\subsection{\bf Static longitudinal solution. Analysis of 
its general structure}
}
\vskip 3mm
\vskip 0.3cm
\noindent
{\small
\subsubsection{\bf Static solution and causal structure}
}
\vskip 3mm

Equations (\ref{MET_11})-(\ref{MAX_0}) are valid for a static 
and rotationally 
symmetric spacetime. One sees that equation (\ref{MET_02}) 
implies that the electric and magnetic fields cannot be 
simultaneously  non-zero, i.e., there is no static dyonic 
solution. In this work we will consider the magnetically 
charged case alone ($A_t=0,\,A_{\varphi} \neq 0$). 
For purely electrically charged solutions of the theory 
see \cite{Zanchin_Lemos}. 
So, assuming vanishing electric field, one has from Maxwell 
equation (\ref{MAX_0}) that 
\begin{equation} 
F^{\varphi r}=\frac{2 \chi_{\rm m}}{r} e^{-\mu},
\label{MAX_1}
\end{equation}
where $\chi_{\rm m}$ is an integration constant which measures
the intensity of the magnetic field source. 
One then has that 
\begin{eqnarray}
T_{rr}=\frac{\chi_{\rm m}^2}{2 \pi \alpha^2 r^2} e^{-2\nu}e^{-2\mu} 
  \:,   \:\:\:\:\:
  T_{\varphi \varphi}=\frac{\chi_{\rm m}^2}{2 \pi \alpha^4 r^2} 
           e^{2\nu}e^{-2\mu}\:.
\label{MAX_2}
\end{eqnarray}
Adding equations (\ref{MET_11}) and (\ref{MET_22}) one obtains 
$\mu_{,rr}=-(\mu_{,r})^2$, yielding for the $g_{zz}$ component 
the solution
\begin{equation}
  e^{2\mu}= \alpha^2 r^2\:,
                                    \label{g_zz}
\end{equation}
where we have defined $\alpha^2\equiv -\Lambda/3>0$. 

The vector potential 
$A=A_{\mu}(r)dx^{\mu}=A_{\varphi}(r)d \varphi$ with 
$A_{\varphi}(r)=\int F_{\varphi r}dr$ is then 
\begin{equation} A=-\frac{2 \chi_{\rm m}}{\alpha^3 r}
    d\varphi \:.
                                    \label{VEC_POTENT}
\end{equation}
Inserting the solutions (\ref{MAX_2}) and (\ref{g_zz}) into 
equation (\ref{MET_11}) we obtain finally the spacetime
generated by the static magnetic source 
\begin{eqnarray} 
ds^2 &=& -(\alpha r)^2 dt^2
         + \frac{dr^2}{(\alpha r)^2
          +b(\alpha r)^{-1} - 4\chi_{\rm m}^2(\alpha^2 r)^{-2}}
                                        \nonumber \\
      & &   + \frac{1}{\alpha^2}{\bigl [}
         (\alpha r)^2 +b(\alpha r)^{-1} 
      - 4\chi_{\rm m}^2(\alpha^2 r)^{-2} {\bigr ]} d\varphi^2
      + (\alpha r)^2 dz^2\:,          
                         \label{Met}  
\end{eqnarray} 
where $b$ is a constant of integration related with the mass
of the solutions, as will be shown.

In order to study the general structure of solution (\ref{Met}),
we first look for curvature singularities. The Kretschmann 
scalar, 
\begin{eqnarray}
 R_{\mu\nu\gamma\sigma}R^{\mu\nu\gamma\sigma} = 
24 \alpha^4 {\biggl [} 1+\frac{b^2}{2(\alpha r)^6}{\biggl ]}
  - \frac{4 \chi_{\rm m}^2}{\alpha^5 r^7}
{\biggl [} b-\frac{7\chi_{\rm m}^2}{6\alpha^3 r}{\biggl ]} \:,  
                                          \label{Kret}
\end{eqnarray}
diverges at $r=0$ and therefore one might think that there is 
a curvature singularity located at $r=0$. However, as will be
seen below, the true spacetime will never achieve $r=0$.

Now, we look for the existence of horizons and, in particular, 
we look for the possible presence of magnetically charged black 
hole solutions. 
We will conclude that there are no horizons and thus no black 
holes.
The horizons, $r_+$, are given by the zeros of 
the $\Delta$ function, where $\Delta=(g_{rr})^{-1}$. 
We find that $\Delta$ has one root $r_+$ given 
by,
\begin{equation}
r_+ = \frac{b^\frac13}{2\alpha} {\biggl [} \sqrt{\frac{2}{\sqrt{s}}-s}-
    \sqrt{s} {\biggr ]}\:,
                             \label{hor_1}
\end{equation}
where
\begin{equation}
s = \left( \frac12 + 
\frac12 \sqrt{1+4\left(\frac{4h^2}{3}\right)^3}\right)^{\frac13} -
 \left(- \frac12 + \frac12 \sqrt{1+4\left(\frac{4h^2}{3}\right)^3}\right)
^{\frac13}\:,
                             \label{horz_2}
\end{equation}
\begin{equation}
h^2 = \frac{16 \chi_{\rm m}^2}{b^\frac43} \:.
                             \label{hor_3}
\end{equation}
The $\Delta(r)$ function is negative for $r<r_+$ and 
positive for $r>r_+$. 
We might then be tempted to say that the solution has an horizon
at $r=r_+$ and consequently that we are in the presence of a magnetically 
charged black hole. However, the above analysis is wrong. 
In fact, we first notice that the metric components 
$g_{rr}=\Delta^{-1}$ and $g_{\varphi \varphi}$ are related 
by $g_{\varphi \varphi}=(\alpha^2 g_{rr})^{-1}$. 
Then, when $g_{rr}$ becomes negative (which occurs for $r<r_+$) 
so does $g_{\varphi \varphi}$ and this leads to an apparent 
change of signature from $+2$ to $-2$. 
This strongly indicates \cite{Hor_Hor} that we are using an incorrect extension 
and that we should choose a different continuation to describe 
the region $r<r_+$. 
Besides, we will verify that one can introduce a new coordinate 
system so that the  spacetime is geodesically complete for 
$r\geq r_+$ \cite{Hor_Hor}. In fact, 
our next step is to show that we can choose a new coordinate 
system for which every null or timelike geodesic starting 
from an arbitrary point either can be extended to infinite 
values of the affine parameter along the geodesic or ends on a 
singularity.

To achieve our aim we introduce the new radial coordinate $\rho$,
\begin{eqnarray}
\rho^2=r^2-r_+^2 \Rightarrow 
    dr^2=\frac{\rho^2}{\rho^2+r_+^2}d \rho^2 \:.
                                       \label{Transf_1}
\end{eqnarray}
With this coordinate change the metric Eq. (\ref{Met})
is written as
\begin{eqnarray}
 \hspace{-1cm} & &  \hspace{-1cm} ds^2  =  
      -\alpha^2 (\rho^2 + r_+^2) dt^2
     + \frac{\frac{\rho^2}{(\rho^2 + r_+^2)}}
     { {\biggl [}  \alpha^2 (\rho^2 + r_+^2) 
       + \frac{b}{[\alpha^2(\rho^2 + r_+^2)]^{1/2}}        
      -\frac{4 \chi_{\rm m}^2} {\alpha^4 (\rho^2 + r_+^2)} 
      {\biggr ]} }          d\rho^2
                                        \nonumber \\
 \hspace{-1cm} & &  \hspace{-1cm} + \frac{1}{\alpha^2}{\biggl [}
        \alpha^2 (\rho^2 + r_+^2) 
       + \frac{b}{[\alpha^2(\rho^2 + r_+^2)]^{1/2}}        
      -\frac{4 \chi_{\rm m}^2} {\alpha^4 (\rho^2 + r_+^2)}
       {\biggr ]} d\varphi^2 +\alpha^2 (\rho^2 + r_+^2) dz^2\:,  
                                            \nonumber \\
       & &                         \label{Met_1}  
\end{eqnarray}
where $0\leq\rho<\infty$ and $0\leq\varphi<2\pi$. The coordinate 
$z$ can have the range $-\infty<z<\infty$, 
or can be compactified,  $0\leq z<2\pi$. 

This spacetime has no horizons and no curvature singularity.
However, it has a conic geometry and in particular it has a 
conical singularity at $\rho=0$. In fact, using a Taylor expansion,
we have that in the vicinity of $\rho=0$ the 
metric Eq. (\ref{Met_1}) is written as 
\begin{eqnarray}
 ds^2 &\sim& -\alpha^2  r_+^2 dt^2 + \frac{1}{\alpha^2 r_+^2}
      \frac{1}{  {\bigl [} 1- (b/2)(\alpha r_+)^{-3}
       +(4\chi_{\rm m}^2/\alpha^2)(\alpha r_+)^{-4}{\bigr ]} }  d\rho^2    
                                   \nonumber \\
     & &   +{\bigl [}  1- (b/2)(\alpha r_+)^{-3}
           +(4\chi_{\rm m}^2/\alpha^2)(\alpha r_+)^{-4}
            {\bigr ]} \rho^2  d\varphi^2 
           + \alpha^2  r_+^2 dz^2\:. 
                                      \label{Met_1_0}  
\end{eqnarray}
Indeed, there is a conical singularity at $\rho=0$ since 
\begin{eqnarray}
\lim_{\rho \rightarrow 0}\frac{1}{\rho}
\sqrt{\frac{g_{\varphi \varphi}}{g_{\rho \rho}}} \neq 1 \:,
                                        \label{Con_Sing}
\end{eqnarray}
i.e, as the radius $\rho$ tends to zero, the limit of the 
ratio ``circumference/radius'' is not $2\pi$.
The conical singularity can be removed if one identifies 
the coordinate $\varphi$ with the period
\begin{eqnarray}
{\rm Period}_{\varphi}=2 \pi {\biggl [}  \lim_{\rho \rightarrow 0}
\frac{1}{\rho} \sqrt{\frac{g_{\varphi \varphi}}
{g_{\rho \rho}}} {\biggr ]}^{-1}=2\pi(1-4\mu) \:,
                                        \label{T_Con}
\end{eqnarray}
with $\mu$ given by
\begin{eqnarray}
\mu= \frac{1}{4} {\biggl [}1- \frac{1} {  
        \alpha r_+ - (b/2)(\alpha r_+)^{-2} 
       +(4\chi_{\rm m}^2/\alpha^2)(\alpha r_+)^{-3} }
 {\biggr ]}\:.
                                        \label{T_Con_1}
\end{eqnarray}
From (\ref{Met_1_0})-(\ref{T_Con_1}) one concludes that 
near the origin, $\rho=0$, metric (\ref{Met_1}) describes 
a spacetime which is locally flat but has a conical
singularity at $\rho=0$ with an angle deficit 
$\delta \varphi=8\pi \mu$. 
Since near the origin our metric (\ref{Met_1_0}) is
identical to the spacetime generated by a cosmic string
we can use the procedure of 
Vilenkin \cite{Vil_string} to show that the stress-energy tensor
of the string is  
\begin{eqnarray}
T_{\mu}^{\nu}=(T_{t}^{t},T_{x}^{x},T_{y}^{y},T_{z}^{z})=
\mu\delta(x)\delta(y)(-1,0,0,-1)\:,
                                        \label{T_String}
\end{eqnarray}
where $\mu$ is the mass per unit length of the string 
defined in (\ref{T_Con_1}) and $(x,y)$ are the
cartesian coordinates, $x=\rho \cos \varphi$ and 
$y=\rho \sin \varphi$.

In (\ref{Met_1}), when one sets $\alpha=0$ and $\chi_{\rm m}=0$ 
one recovers the Levi-Cevita solution \cite{L-C}.
If one sets  $\alpha=0$ one recovers the Witten solution 
\cite{Witten} (see also \cite{Kramers}). So the present paper is 
an extension to include the cosmological constant, rotation 
and the definition of conserved quantities (see below).

Note also that there are two main distinct properties 
relatively to the electric charged solutions 
\cite{Zanchin_Lemos}. First, the 
electric solutions have black holes, while the magnetic 
do not. Second, the electric solutions can have cylindrical, 
toroidal and planar topologies for the 2-space generated 
by the Killing vectors $\partial_\varphi$ and $\partial_z$, 
whereas the magnetic solutions can only have cylindrical and 
toroidal topologies, the first case represents an infinite 
straight magnetic string and the second a closed one.

\vskip 0.3cm
\noindent
{\small
\subsubsection{\bf Geodesic structure}
}
\vskip 3mm

We want to show that the spacetime described by  (\ref{Met_1}) 
is both null and timelike geodesically complete. 
The equations governing the geodesics can be derived from the
Lagrangian
\begin{equation}
{\cal{L}}=\frac{1}{2}g_{\mu\nu}\frac{dx^{\mu}}{d \tau}
       \frac{dx^{\nu}}{d \tau}=-\frac{\delta}{2}\:,
                                 \label{LAG)}  
\end{equation}
where $\tau$ is an affine parameter along the geodesic which, 
for a timelike geodesic, can be identified with the proper 
time of the particle along the geodesic. For a null geodesic 
one has $\delta=0$ and for a timelike geodesic $\delta=+1$. 
From the Euler-Lagrange equations one gets that the generalized 
momentums associated with the time coordinate, angular 
coordinate and $z$-component are constants: 
$p_t=E$, $p_{\varphi}=L$, $p_z=P$. The constant $E$ is related 
to the timelike Killing vector $(\partial/\partial t)^{\mu}$ 
which reflects the time translation invariance of the metric, 
while the constant $L$ is associated to the spacelike Killing 
vector $(\partial/\partial \varphi)^{\mu}$ which reflects the 
invariance of the metric under rotation and the constant $P$ 
is associated to the spacelike Killing vector 
$(\partial/\partial z)^{\mu}$ which reflects the 
invariance of the metric under a boost in the $z$ direction. 
Note that since the spacetime is not asymptotically flat, 
the constants $E$, $L$ and $P$ cannot be interpreted as 
the energy, angular momentum and linear momentum at 
infinity. 

From the metric we can derive directly the radial geodesic, 
\begin{eqnarray}
\dot{\rho}^2=-\frac{1}{g_{\rho\rho}}
 \frac{E^2 g_{\varphi \varphi}+L^2 g_{tt}}
       {g_{tt} g_{\varphi \varphi} } 
      -\frac{P^2}{g_{\rho\rho} g_{zz}} 
        -\frac{\delta}{g_{\rho\rho}} \:. 
                                        \label{GEOD_1}
\end{eqnarray}
Now, using the two useful relations 
$g_{\varphi \varphi}=\rho^2[\alpha^2(\rho^2+r_+^2)
g_{\rho\rho}]^{-1}$ and 
$g_{tt} g_{\varphi \varphi}=-\rho^2/g_{\rho\rho}$, 
we can write Eq. (\ref{GEOD_1}) as 
\begin{eqnarray}
\rho^2 \dot{\rho}^2= \frac{\rho^2}{g_{\rho\rho}}{\biggl [}
 \frac{E^2-P^2}{\alpha^2 (\rho^2+r_+^2)}-\delta {\biggr ]} 
       -L^2 \alpha^2 \rho^2 \:. 
                                     \label{GEOD_2}
\end{eqnarray}
Noticing that $1/g_{\rho\rho}$ is always positive for $\rho>0$ and
zero for $\rho=0$, we conclude 
the following about the null geodesic motion ($\delta=0$). 
(i) if $E^2>P^2$, 
null spiraling ($L\neq 0$) particles that start at 
$\rho=+\infty$ spiral toward a turning point 
$\rho_{\rm tp}>0$ and then return back to infinity,
(ii) the null particle coming from infinity hits the 
conical singularity with vanishing velocity if and 
only if $E^2>P^2$ and $L=0$, (iii) whatever the value of
$L$ is, for $E^2<P^2$ there is no possible null geodesic
motion, (iv) the same occurs if $E=P$ and $L\neq 0$,
(v) if $E=P$ and $L=0$, whatever the value of $\rho$ is, 
one has $\dot{\rho}=0$ and 
$\dot{\varphi}=-Lg_{tt}g_{\rho\rho}/\rho^2=0$ so the null
particle moves in a straight line along the $z$ direction.

Now, we analyse the timelike geodesics ($\delta=+1$).
Timelike geodesic motion is possible only if the energy and 
linear momentum of the 
particle satisfies $E^2-P^2 > \alpha^2 r_+^2$. 
In this case, spiraling ($L\neq 0$) timelike particles are bounded 
between two turning points, a and b, that satisfy 
$\rho_{\rm tp}^{\rm a} > 0$ and 
$\rho_{\rm tp}^{\rm b} < \sqrt{(E^2-P^2)/\alpha^2 - r_+^2}$, 
with $\rho_{\rm tp}^{\rm b} \geq \rho_{\rm tp}^{\rm a}$. 
When the timelike particle has no angular momentum ($L=0$) 
there is a turning point located at 
$\rho_{\rm tp}^{\rm b}=\sqrt{(E^2-P^2)/\alpha^2 - r_+^2}$ 
and it hits the conical singularity at $\rho=0$.

Hence, we confirm that the spacetime described by 
Eq. (\ref{Met_1}) is both timelike and null geodesically 
complete.

\vskip 1cm
\noindent
{\small
\subsection{\bf The general rotating longitudinal solution}
}
\vskip 3mm
Now, we want to endow our spacetime solution with a global
rotation, i.e we want to add angular momentum to the spacetime.
In order to do so we perform the following rotation boost in 
the $t$-$\varphi$ plane (see e.g. 
\cite{Zanchin_Lemos,Sa_Lemos_Static,OscarLemos,HorWel})
\begin{eqnarray}
 t &\mapsto& \gamma t-\frac{\omega}{\alpha^2} \varphi \:,
                                       \nonumber  \\
 \varphi &\mapsto& \gamma \varphi-\omega t \:,
                                       \label{TRANSF_J}
\end{eqnarray}
where $\gamma$ and $\omega$ are constant parameters. 
Substituting (\ref{TRANSF_J}) into (\ref{Met_1}) we obtain 
\begin{eqnarray} 
ds^2 \hspace{-0.2cm} &=& \hspace{-0.2cm}  
    -{\biggl [} \frac{\alpha^2 (\rho^2 + r_+^2)} 
      {(\gamma^2-\omega^2 / \alpha^2)^{-1}}
     -\frac{\omega^2}{\alpha^2}\frac{b}
    {[\alpha^2 (\rho^2 + r_+^2)]^{\frac{1}{2}}}
    +\frac{\omega^2}{\alpha^2}\frac{4 \chi_{\rm m}^2}
    {\alpha^4 (\rho^2 + r_+^2)} {\biggr ]} dt^2   
                                          \nonumber \\
      & &      -\frac{\gamma \omega}{\alpha^2}
  {\biggl [}b [\alpha^2 (\rho^2 + r_+^2)]^{-\frac{1}{2}}
   - 4\chi_{\rm m}^2 [\alpha^4 (\rho^2 + r_+^2)]^{-1}  
        {\biggr ]} 2dt d\varphi   
                                         \nonumber \\
    & &     + \frac{\frac{\rho^2}{(\rho^2 + r_+^2)}}
       { {\biggl [}  \alpha^2 (\rho^2 + r_+^2) + 
   \frac{b}{[\alpha^2 (\rho^2 + r_+^2)]^{\frac{1}{2}}}
     -\frac{4\chi_{\rm m}^2}
   {\alpha^4 (\rho^2 + r_+^2)} {\biggr ]}}   d\rho^2
                                             \nonumber \\
   & &   + \frac{1}{\alpha^2}{\biggl [}
   \frac{\alpha^2 (\rho^2 + r_+^2)}
    {(\gamma^2-\omega^2 / \alpha^2)^{-1}} 
    +\frac{\gamma^2 b}
   {[\alpha^2 (\rho^2 + r_+^2)]^{\frac{1}{2}}}
    -\frac{4 \gamma^2  \chi_{\rm m}^2}
   {\alpha^4 (\rho^2 + r_+^2)} {\biggr ]}   d\varphi^2 
                                       \nonumber \\   
                          & & +\alpha^2(\rho^2+r_+^2) dz^2 \:.    
                           \label{Met_1_J}  
\end{eqnarray}
Inserting transformations (\ref{TRANSF_J}) into 
(\ref{VEC_POTENT}) we obtain that the vector potential 
$A=A_{\mu}(\rho)dx^{\mu}$ is now given by
\begin{equation}
A=-\omega A(\rho)dt +\gamma  A(\rho) d\varphi\:,
                                    \label{VEC_POTENT_J}
\end{equation}
where 
$A(\rho)=-[2\chi_{\rm m}/(\alpha^3 r_+)] 
\ln[2(r_+ + \sqrt{\rho^2+r_+^2})/\rho]$.

We choose $\gamma^2-\omega^2 / \alpha^2=1$ because in this way  
when the angular momentum vanishes ($\omega=0$) we have 
$\gamma=1$ and so we recover the static solution. 
Solution (\ref{Met_1_J}) represents a magnetically charged 
cylindrical stationary spacetime (a spinning magnetic string) 
and also solves (\ref{ACCAO}). 
Transformations (\ref{TRANSF_J}) generate a new metric 
because they are not permitted global coordinate transformations 
\cite{Stachel}. Transformations (\ref{TRANSF_J}) can be done 
locally, but not globally. Therefore, the metrics (\ref{Met_1})
and (\ref{Met_1_J}) can be locally mapped into each other but 
not globally, and as such they are distinct.

\vskip 1cm
\noindent
{\small
\subsection{\bf  Mass, angular momentum and electric charge of 
the longitudinal solution}
}
\vskip 3mm

As we shall see the spacetime solutions (\ref{Met_1})
and (\ref{Met_1_J}) are asymptotically anti-de Sitter.
This fact allows us to calculate the mass, angular momentum 
and the electric charge of the static and rotating solutions. 
To obtain these 
quantities  we apply the formalism  of Regge and Teitelboim 
\cite{Regge} (see also 
\cite{Sa_Lemos_Static,OscarLemos,HorWel,BTZ_Q}).
We first write the metric (\ref{Met_1}) in the canonical form 
involving the lapse function $N^0(\rho)$ and the shift 
function $N^{\varphi}(\rho)$
\begin{equation}
     ds^2 = - (N^0)^2 dt^2
            + \frac{d\rho^2}{f^2}
            + H^2(d\varphi+N^{\varphi}dt)^2 +W^2dz^2\:,
                               \label{MET_CANON}
\end{equation}
where $f^{-2}=g_{\rho\rho}$, $H^2=g_{\varphi \varphi}$, 
$W^2=g_{zz}$, $H^2 N^{\varphi}=g_{t \varphi}$ and 
$(N^0)^2-H^2(N^{\varphi})^2=g_{tt}$.
Then, the action can be written in the Hamiltonian form as a 
function of the energy constraint ${\cal{H}}$, momentum constraint 
${\cal{H}}_{\varphi}$ and Gauss constraint $G$
\begin{eqnarray}
S &=& -\int dt d^3x[N^0 {\cal{H}}+N^{\varphi} {\cal{H}_{\varphi}}
      + A_{t} G]+  {\cal{B}}          \nonumber \\
 &=&  -\Delta t \int d\rho N \frac{\Delta z}{8}
        {\biggl [} \frac{128 \pi^2}{H^3 W}
        + (f^2)_{,\rho}(HW)_{,\rho}
        +2f^2(H_{,\rho}W)_{,\rho}+2f^2HW_{,\rho\rho}                   
             \nonumber \\
 & &    -2\Lambda HW+ \frac{2HW}{f}(E^2+B^2){\biggr ]}
                                        \nonumber \\
 & &   + \Delta t \int d\rho N^{\varphi} \frac{\Delta z}{8}
        {\biggl [}{\bigl (}2 \pi 
       {\bigr )}_{,\rho}+\frac{4HW}{f}E^{\rho}B{\biggr ]} 
\nonumber \\
 & &    + \Delta t \int d\rho A_t \frac{\Delta z}{8}
        {\biggl [}-\frac{4HW}{f}
        \partial_{\rho} E^{\rho}{\biggr ]} +{\cal{B}} \:, 
                              \label{ACCAO_CANON}
\end{eqnarray}
where $N=N^0/f$, 
$\pi \equiv {\pi_{\varphi}}^{\rho}=
-\frac{fH^3W(N^{\varphi})_{,\rho}}{2N^0}$ 
(with $\pi^{\rho \varphi}$ being the momentum conjugate to 
$g_{\rho \varphi}$),  $E^{\rho}$ and $B$ are the electric and 
magnetic fields 
and ${\cal{B}}$ is a boundary term.
Upon varying the action with respect to $f(\rho)$, $H(\rho)$, 
$W(\rho)$, $\pi(\rho)$, and $E^{\rho}(\rho)$ one picks up 
additional surface terms.
Indeed,
\begin{eqnarray}
\delta S &=& - \Delta t N \frac{\Delta z}{8}
         {\biggl [}(HW)_{,\rho} \delta f^2 
           -(f^2)_{,\rho}W\delta H 
         +2f^2 W \delta (H_{,\rho})                            
                                                   \nonumber \\
         & &              
         -H(f^2)_{,\rho}\delta W+ 2Hf^2\delta (W_{,\rho}) 
         {\biggr ]}
                                        \nonumber \\
         & & +\Delta t N^{\varphi}\frac{\Delta z}{4}\delta \pi+ 
        \Delta t A_t \frac{\Delta z}{8}
          {\biggl [}- \frac{4HW}{f} \delta E^{\rho}{\biggr ]}         
+ \delta {\cal{B}}
          \nonumber \\
         & & +(\mbox{terms vanishing when the
                    equations of motion hold}).
                               \label{DELTA_ACCAO}
\end{eqnarray}
In order that the Hamilton's equations are satisfied,
the boundary term ${\cal{B}}$ has to be adjusted so that it cancels 
the above additional surface terms. More specifically one has
\begin{equation}
  \delta {\cal{B}} = -\Delta t N \delta \bar{M} 
          +\Delta t N^{\varphi}\delta \bar{J}+
             \Delta t A_t \delta \bar{Q}_{\rm e} \:,
                              \label{DELTA_B}
\end{equation}
where one identifies $\bar{M}$ as the mass, $\bar{J}$ as 
the angular momentum and $\bar{Q}_{\rm e}$ as the electric 
charge since they are the terms conjugate to the 
asymptotic values of $N$, $N^{\varphi}$ and $A_t$, respectively.

To determine the mass, angular momentum and the electric 
charge of the solutions one must take the spacetime that 
we have obtained and subtract the background reference 
spacetime contribution, i.e., we choose the energy zero 
point in such a way that the mass, angular momentum and 
charge vanish when the matter is not present.

Now, note that spacetime (\ref{Met_1_J}) has an asymptotic 
metric given by 
\begin{equation}
-{\biggl (}\gamma^2-\frac{\omega^2}{\alpha^2} {\biggr )} 
    \alpha^2 \rho^2 dt^2+ \frac{d \rho^2}{ \alpha^2 \rho^2}
+ {\biggl (}\gamma^2-\frac{\omega^2}{\alpha^2} {\biggr )} 
\rho^2 d \varphi^2 +\alpha^2 \rho^2 dz^2\:,
                                         \label{ANTI_SITTER}
\end{equation}
where $\gamma^2-\omega^2 / \alpha^2=1$ so, it is asymptotically 
a cylindrical anti-de Sitter spacetime (we can rescale the coordinates $r$ 
and $z$ so that the usual form of the anti-de Sitter spacetime 
becomes apparent). 
The cylindrical anti-de Sitter spacetime is 
also the background reference spacetime, since the metric 
(\ref{Met_1_J}) reduces to (\ref{ANTI_SITTER}) if the 
matter is not present ($b=0$ and $\chi_{\rm m}=0$).
Taking the subtraction of the background reference spacetime 
into account and noting that $W-W_{\rm ref}=0$ and that
$W_{,\rho}-W_{,\rho}^{\rm ref}=0$ we have that the mass,
angular momentum and electric charge are given by
\begin{eqnarray}
\bar{M} \hspace{-0.3 cm}&=& \hspace{-0.3 cm}\frac{\Delta z}{8}{\biggl [}      
      -(HW)_{,\rho}(f^2-f^2_{\rm ref})      
      +(f^2)_{,\rho}W(H-H_{\rm ref}) 
     -2f^2W(H_{,\rho}-H_{,\rho}^{\rm ref}){\biggr ]}\:, 
                                      \nonumber \\ 
 & &                                \label{Massa}  \\ 
\bar{J} \hspace{-0.3 cm} &=& \hspace{-0.3 cm}
 - \Delta z(\pi-\pi_{\rm ref})/4 \:,
                                   \label{Angular} \\
\bar{Q}_{\rm e} \hspace{-0.3 cm} &=& \hspace{-0.3 cm}
           \frac{\Delta z}{2} \frac{HW}{f}
            (E^{\rho}-E^{\rho}_{\rm ref}) \:.         
                      \label{Carga}
\end{eqnarray}  
Then, we finally have that the mass per unit length 
($\bar{M}/\Delta z$) and the angular momentum per unit 
length ($\bar{J}/\Delta z$) are (after taking the 
asymptotic limit, $\rho \rightarrow +\infty$)
\begin{eqnarray}
M &=& \frac{b}{8}{\biggl [}
       \gamma^2+2\frac{\omega^2}{\alpha^2}{\biggr ]} 
         + {\rm Div_M}(\chi_{\rm m},\rho) 
       =M_0  + {\rm Div_M}(\chi_{\rm m},\rho) \:,
                        \label{M} \\  
J &=& \frac{3b}{8}\frac{\gamma \omega}{\alpha^2}
      + {\rm Div_J}(\chi_{\rm m},\rho) \:,
                          \label{J}
\end{eqnarray}  
where ${\rm Div_M}(\chi_{\rm m},\rho)$ and 
${\rm Div_M}(\chi_{\rm m},\rho)$ are terms proportional 
to the magnetic source $\chi_{\rm m}$ that diverge as 
$\rho \rightarrow +\infty$. The presence of these kind 
of  divergences in the mass is a usual feature present 
in charged solutions. They can be found for example on 
the electrically charged point source solution in
3D gravity \cite{Deser_Maz}, in the electrically charged 
BTZ black hole \cite{BTZ_Q} and in the electrically 
charged solutions of three-dimensional Brans-Dicke 
action \cite{OscarLemos}. Following 
\cite{Deser_Maz,BTZ_Q} (see also \cite{OscarLemos}) 
these divergences can be treated as follows. One 
considers a boundary of large radius $\rho_0$ 
involving the system. Then, one sums and subtracts 
${\rm Div_M}(\chi_{\rm m},\rho_0)$ to (\ref{M}) 
so that the mass per unit length (\ref{M}) is now written as 
\begin{equation}
M = M(\rho_0)+ [{\rm Div_M}(\chi_{\rm m},\rho)
     -{\rm Div_M}(\chi_{\rm m},\rho_0)] \:,
          \label{M0_0}
\end{equation}  
where $M(\rho_0)=M_0+{\rm Div_M}(\chi_{\rm m},\rho_0)$, i.e., 
\begin{equation}
M_0=M(\rho_0)-{\rm Div_M}(\chi_{\rm m},\rho_0)\:.   
                    \label{M0_0_v2}
\end{equation}  
The term between brackets in (\ref{M0_0}) vanishes when 
$\rho \rightarrow \rho_0$. Then $M(\rho_0)$ is the 
energy within the radius $\rho_0$. The difference 
between $M(\rho_0)$ and $-M_0$ is 
$-{\rm Div_M}(\chi_{\rm m},\rho_0)$ which is 
interpreted as the electromagnetic energy outside 
$\rho_0$ apart from an infinite constant which is 
absorbed in $M(\rho_0)$. The sum (\ref{M0_0_v2}) is 
then independent of $\rho_0$, finite and equal to the 
total mass.
In practice the treatment of the mass divergence 
amounts to forgetting about $\rho_0$ and take as 
zero the asymptotic limit: 
$\lim {\rm Div_M}(\chi_{\rm m},\rho)=0$. 

To handle the angular momentum divergence, one first 
notices that the asymptotic limit of the angular 
momentum per unit  mass $(J/M)$ is either zero or one, 
so the angular momentum diverges at a rate slower or 
equal to the rate of the mass divergence. 
The divergence on the angular momentum can then be 
treated in a similar way as the mass divergence. 
So, one can again consider a boundary of large 
radius $\rho_0$ involving the system. Following the 
procedure applied for the mass divergence one 
concludes that the divergent term 
$-{\rm Div_J}(\chi_{\rm m},\rho_0)$ can be interpreted 
as the electromagnetic angular momentum outside  
$\rho_0$ up to an infinite constant that is absorbed 
in $J(\rho_0)$. 

Hence, in practice the treatment of both the mass and 
angular divergences amounts to forgetting about $\rho_0$ 
and take as zero the asymptotic limits: 
$\lim {\rm Div_M}(\chi_{\rm m},\rho)=0$ and 
$\lim {\rm Div_J}(\chi_{\rm m},\rho)=0$ in 
Eqs. (\ref{M})-(\ref{J}). 

Now, we calculate the electric charge of the solutions.
To determine the electric field we must consider the 
projections of the Maxwell field on spatial hypersurfaces. 
The normal to such hypersurfaces is 
$n^{\nu}=(1/N^0,0,-N^{\varphi}/N^0,0)$ and the electric 
field is given by 
$E^{\mu}=g^{\mu \sigma}F_{\sigma \nu}n^{\nu}$. Then, from 
(\ref{Carga}), the electric charge per unit length 
($\bar{Q}_{\rm e}/\Delta z$) is 
\begin{equation} 
Q_{\rm e}=-\frac{4HWf}{N^0}(\partial_{\rho}A_t
-N^{\varphi} \partial_{\rho} A_{\varphi})
=\frac{\omega}{\alpha^2}   \chi_{\rm m} \:.
\label{Q}
\end{equation}
Note that the electric charge is proportional
to $\omega \chi_{\rm m}$.
In the next subsection we will propose a physical 
interpretation for the origin of the magnetic field 
source and discuss the result obtained in (\ref{Q}).

Now, we want to cast the metric (\ref{Met_1_J}) in terms 
of $M$, $J$ and $Q_{\rm e}$. We can use (\ref{M}) and 
(\ref{J}) to solve a quadratic equation for $\gamma^2$ 
and $\omega^2 / \alpha^2$. It gives two distinct sets 
of solutions
\begin{equation}
\gamma^2=\frac{4M}{b}(2- \Omega) \:,\:\:\:\:\:\:\:
  \frac{\omega^2}{\alpha^2}= \frac{2M}{b}\Omega\:, 
                               \label{DUAS}
\end{equation}
\begin{equation}
\gamma^2=\frac{4M}{b}\Omega \:,\:\:\:\:\:\:\: 
\frac{\omega^2}{\alpha^2}= \frac{2M}{b}(2- \Omega)\:, 
\label{DUAS_ERR}
\end{equation}
where we have defined a rotating parameter $\Omega$ as 
\begin{equation} 
\Omega \equiv 1- \sqrt{1-\frac{8}{9}\frac{J^2 \alpha^2}{M^2}}\:. 
                   \label{OMEGA}
\end{equation}
When we take $J=0$ (which implies $\Omega=0$), (\ref{DUAS}) 
gives $\gamma \neq 0$ and $\omega= 0$ while 
(\ref{DUAS_ERR}) gives the nonphysical solution $\gamma=0$ 
and $\omega \neq 0$ which does not reduce to the static 
original metric. Therefore we will study the solutions 
found from (\ref{DUAS}).

The condition that $\Omega$ remains real imposes a 
restriction on the allowed values of the angular momentum:
\begin{equation}
\alpha^2 J^2 \leq \frac{8}{9}M^2 \:.
                    \label{Rest_OMEGA}
\end{equation}
The parameter $\Omega$ ranges between 
$0 \leq \Omega \leq 1$. The condition 
$\gamma^2-\omega^2/\alpha^2=1$ fixes the value of $b$ as a 
function of $M,\Omega$,
\begin{eqnarray} 
b &=& 2M (4-3\Omega) \:.
                                \label{b} 
\end{eqnarray}
The metric (\ref{Met_1_J}) may now be cast in the form
{\small 
\begin{eqnarray}
\hspace{-1 cm} & & \hspace{-0.5cm} ds^2 = \nonumber \\
     & & \hspace{-0.5cm}    
   -{\biggl [}\alpha^2 (\rho^2 + r_+^2)-2M\Omega 
   [\alpha^2 (\rho^2 + r_+^2)]^{-1/2} +4Q^2_{\rm e}
     [\alpha^2 (\rho^2 + r_+^2)]^{-1} {\biggr ]} dt^2
                                             \nonumber \\
    & &   \hspace{-0.5cm}   
         -\frac{8}{3}J{\biggl [}[\alpha^2 
        (\rho^2 + r_+^2)]^{-1/2}  -\frac{2Q^2_{\rm e}}
        {M \Omega}[\alpha^2 (\rho^2 + r_+^2)]^{-1}  
        {\biggr ]} 2dt d\varphi  
                                                 \nonumber \\
    & &     \hspace{-0.5cm} 
       + \frac{\frac{\rho^2}{(\rho^2 + r_+^2)}}
    { {\biggl [}  \alpha^2 (\rho^2 + r_+^2) +2M (4-3\Omega)     
     [\alpha^2 (\rho^2 + r_+^2)]^{-1/2}
     -4\chi_{\rm m}^2 [\alpha^4 (\rho^2 + r_+^2)]^{-1} 
     {\biggr ]}}           d\rho^2
                                             \nonumber \\
    & &   \hspace{-0.5cm}
      + \frac{1}{\alpha^2}{\biggl [}\alpha^2 (\rho^2 + r_+^2)  
       +\frac{4M(2-\Omega)}{[\alpha^2 (\rho^2 + r_+^2)]^{1/2}}
    -\frac{8(2-\Omega)}{4-3\Omega}\frac{\chi_{\rm m}^2}   
     {[\alpha^4 (\rho^2 + r_+^2)]}      
         {\biggr ]}   d\varphi^2  \nonumber \\
    & &  \hspace{-0.5cm} + \alpha^2(\rho^2 + r_+^2) dz^2 \:.
                           \label{Met_MJQ}  
\end{eqnarray}
}
The static solution can be obtained by putting 
$\Omega=0$ (and thus $J=0$ and $Q_{\rm e}=0$) on the above expression 
[see (\ref{TRANSF_J})].

\vskip 3cm
\noindent
{\small
\subsection{\bf Physical interpretation of the longitudinal
 magnetic field source}
}
\vskip 3mm

When we look back to the value of the electric charge per
unit length, Eq. (\ref{Q}), we see that it is zero when 
the angular momentum of the spacetime vanishes, $J=0=\omega$.  
This result, no electric source in the static spacetime case,
was expected since 
we have imposed that the static electric field is zero ($F_{21}$ is 
the only non-zero component of the Maxwell tensor). 

Still missing however, is a physical interpretation for the 
origin of the magnetic field source.
It is quite evident that 
the magnetic field source is not a 't Hooft-Polyakov monopole 
since we are working with the Maxwell theory and not with an 
$SO(3)$ gauge theory spontaneously broken to $U(1)$ by an 
isotriplet Higgs field. We might then think that the magnetic 
field is produced by a Dirac line-like monopole. However, this 
is not also the case since the Dirac monopole with strength 
$g_{\rm m}$ appears when one breaks the Bianchi identity 
\cite{Dirac}: 
$\partial_{\nu} (\sqrt{-g} \tilde{F}^{\mu\nu}) = 
4\pi k^{\mu}/\sqrt{-g}$, 
where 
$\tilde{F}^{\mu \nu}=\epsilon^{\mu \nu \gamma \sigma}
F_{\gamma \sigma}/2$ is the dual of the Maxwell field strength
and 
$k^{\mu}=\sum g_{\rm m} \delta^3(\vec{x}-\vec{x}_0)\dot{x}^{\mu}$
is the magnetic current density.
But in this work we are clearly dealing with the Maxwell theory 
which satisfies Maxwell equations and the  Bianchi identity 
\begin{eqnarray}
& &  \frac{1}{\sqrt{-g}}\partial_{\nu}(\sqrt{-g}F^{\mu\nu})=
    {4\pi}\frac{1}{\sqrt{-g}} j^{\mu} \:,                
                                \label{Max_j} \\
& &  \partial_{\mu} 
     (\sqrt{-g} \tilde{F}^{\mu\nu} )=0 \:,         
                                    \label{Max_bianchi}
\end{eqnarray}  
respectively. We have made use of the fact that the general 
relativistic current density is $1/\sqrt{-g}$ times the 
special relativistic current density 
$j^{\mu}=\sum q \delta^3(\vec{x}-\vec{x}_0)\dot{x}^{\mu}$. 
Hence, from (\ref{Max_bianchi}) we have to put away the 
hypothesis of the Dirac monopole.  

Following \cite{Witten} the magnetic field source can then 
be interpreted as
composed by a system of two symmetric and superposed  
electrically charged lines along the $z$ direction 
(each with strength $q$ per unit length). One of the electrically 
charged lines is at rest and the other is spinning around the 
$z$ direction with an angular velocity $\dot{\varphi}_0$. 
Clearly, this system 
produces no electric field since the total electric charge is 
zero and the magnetic field is produced by the rotating 
electric current. 
To confirm our interpretation, we go back to Eq. (\ref{Max_j}). 
In our solution, the only non-vanishing component of the 
Maxwell field is $F^{\varphi \rho}$ which implies that only 
$j^{\varphi}$ is not zero. According to our interpretation 
one has $j^{\varphi}=q \delta^2(\vec{x}-\vec{x}_0)
\dot{\varphi}$, 
which one inserts in Eq. (\ref{Max_j}). Finally, integrating 
over $\rho$ and  $\varphi$  and introducing 
$F^{\varphi \rho}=2 \chi_{\rm m}/(\alpha \rho \sqrt{\rho^2+r_+^2})$ 
we have 
\begin{equation} 
\chi_{\rm m} \propto q \dot{\varphi}_0 \:.
      \label{Q_mag}
\end{equation} 
So, the magnetic source strength per unit length, 
$\chi_{\rm m}$, can be interpreted as an electric charge 
per unit length times its spinning velocity.

Looking again to the value of the electric charge per
unit length, Eq. (\ref{Q}), one sees that after applying the 
rotation boost in the $t$-$\varphi$ plane to endow our 
initial static spacetime with angular momentum, there 
appears  a net electric charge. This result was once again 
expected since now, besides the magnetic field along
the $z$ direction
($F_{\rho \varphi} \neq 0$), there is also a radial 
electric field ($F_{t \rho} \neq 0$). A physical 
interpretation for the appearance of the net electric 
charge is needed. To do so, we first go back to the static 
spacetime. According to our interpretation, an 
observer at rest relative to the source ($S$) sees a 
density of charges at rest 
which is equal to the negative of the density of charges 
that are rotating. Now, we consider the stationary 
spacetime and we look to the point of view of an 
observer ($S'$) that follows the intrinsic rotation of 
the spacetime, i.e. that rotates with the same angular 
velocity of the spacetime. The observer $S'$ is moving 
relatively to the observer $S$ and we know that the charge 
density of a moving distribution of charges varies as the 
inverse of the relativistic length (since the density is 
a charge over an area) when we compare the measurements 
made by two different frames. So, the two set of charge 
distributions that had symmetric charge densities in the 
frame $S$ will not have charge densities with equal 
magnitude in the frame $S'$. Hence, they will not cancel 
each other in the frame $S'$ and a net electric charge 
appears. This analysis is similar to the one that occurs 
when one has a cupper wire with an electric current along
the wire and we apply a translation Lorentz boost to the 
wire along its direction: initially, 
there is only a magnetic field  but, after the Lorentz 
boost, one also has an electric field. The only 
difference is that in the present situation the Lorentz 
boost is a rotational one and not a translational one.

\vskip 1cm 
\noindent
{\small
\section{\bf Angular magnetic field solution}
}
\vskip 3mm

In section 2 we have found a spacetime generated by a 
magnetic source that produces a longitudinal magnetic field 
along the $z$ direction. In this section we find a spacetime
generated by a magnetic source that produces an angular 
magnetic field along the $\varphi$ direction. Following the steps
of section 2 but now with the roles of  $\varphi$ and $z$
interchanged, we
can write directly the metric and electric potential for the angular
magnetic field solution as
\begin{eqnarray}
 \hspace{-1cm} & & \hspace{-1cm} ds^2 = 
 -\alpha^2 (\rho^2 + r_+^2)dt^2 
  + \frac{\frac{\rho^2}{(\rho^2 + r_+^2)}}
      { {\biggl [}\alpha^2 (\rho^2 + r_+^2)  
     + \frac{b}{[\alpha^2(\rho^2 + r_+^2)]^{1/2}}  
     -\frac{4 \chi_{\rm m}^2} {\alpha^2 (\rho^2 + r_+^2)} 
      {\biggr ]} } d\rho^2
                                   \nonumber \\
 \hspace{-1cm} & & \hspace{-1cm} + (\rho^2 + r_+^2) d\varphi^2 + 
{\biggl [}\alpha^2
(\rho^2 + r_+^2) + \frac{b}{[\alpha^2(\rho^2 + r_+^2)]^{1/2}} 
-\frac{4 \chi_{\rm m}^2} {\alpha^2 (\rho^2 + r_+^2)}
 {\biggr ]}
dz^2\:,                           \nonumber \\
                               & & \label{Met_11} 
\end{eqnarray}
where again $\alpha^2 \equiv -\Lambda/3>0$, $\rho$ is a radial
coordinate, $\varphi$ an angular coordinate
and $z$ ranges between $-\infty<z<\infty$ (or $0\leq z<2\pi$),
\begin{equation}
A=-\frac{2 \chi_{\rm m}}{\alpha^3 r_+} 
\ln {\biggl [}\frac{2(r_+ + \sqrt{\rho^2+r_+^2})}{\rho}  
{\biggr ]} dz \:.
                                    \label{VEC_POTENT_P0}
\end{equation}
This spacetime has a mass per unit lenght given by $M=b/8$
and no electric charge.
The Kretschmann scalar does not diverge for any $\rho$ so 
there is no curvature singularity. Spacetime  (\ref{Met_11})
is also free of conical  singularities.
Besides, if one studies the radial geodesic motion we conclude that
these geodesics can pass through $\rho=0$ 
(which is free of singulaties) from positive values to negative 
values of $\rho$. This shows that in (\ref{Met_11}) the radial
coordinate can take the values $-\infty<\rho<\infty$.
This analysis might suggest that we are in the presence of
a traversable (since the spacetime has no horizons) wormhole 
with a throat of dimension $r_+$.
However, in the vicinity of $\rho=0$,  metric (\ref{Met_11}) 
is written as 
\begin{eqnarray}
 ds^2 &\sim& -\alpha^2  r_+^2 dt^2 + \frac{1}{\alpha^2 r_+^2}
      \frac{1}{  {\bigl [} 1- (b/2)(\alpha r_+)^{-3}
       +4\chi_{\rm m}^2(\alpha r_+)^{-4}{\bigr ]} }  d\rho^2    
                                   \nonumber \\
     & &   +  r_+^2   d\varphi^2 
           +{\bigl [}  1- (b/2)(\alpha r_+)^{-3}
           +4\chi_{\rm m}^2 (\alpha r_+)^{-4}
            {\bigr ]} \rho^2 dz^2\:. 
                                      \label{Met_1_1}  
\end{eqnarray}
which clearly shows that at $\rho=0$ the $z$-direction
collapses and therefore we have to abandon the 
wormhole interpretation.

A physical interpretation for the source of the angular magnetic field
can be given.  The spacetime has zero electric field and angular
magnetic field along the $\varphi$ direction.  The source for the
magnetic field could then be interpreted as composed by a system of
two symmetric and superposed electrically charged lines along the $z$
direction.  One of the electrically charged lines would be at rest and
the other would have a velocity along the $z$ direction. Clearly, this
system produces no electric field since the total electric charge is
zero and the magnetic field is produced by the axial electric current.
There is however a great problem with this interpretation if we want
to extend the solution down to the origin.  Indeed, at $\rho=0$ there
is no physical object to support the proposed source, and in addition,
the four-dimensional character of the solution is lost (since near the
origin the $z$ direction collapses).  In order to have a full solution
one has to consider a matter source up to a boundary radius $\rho_{\rm
b}$ with an appropriate stress-energy tensor which should generate the
exterior solution (\ref{Met_11}).

To add linear momentum to the spacetime, we perform the 
following translation boost in the $t$-$z$ plane: 
$t \mapsto \gamma t-\frac{\lambda}{\alpha} z\,$,
$z \mapsto \gamma z-\frac{\lambda}{\alpha} t\,$,
where $\gamma$ and $\lambda$ are constant parameters
that are choosen so that $\gamma^2-\lambda^2 / \alpha^2=1$ 
because in this way  
when the linear momentum vanishes ($\lambda=0$) we have 
$\gamma=1$ and so we recover the static solution. 
Contrarily to transformation (\ref{TRANSF_J}), the 
above translational boost transformation is permitted since 
$z$ is not an angular coordinate. Thus boosting the solution 
(\ref{Met_11}) through this permitted transformation does not
yield a new solution. However, it generates an electric
field. For the sake of completeness we write below the
boosted solution:
{\small 
\begin{eqnarray}
\hspace{-1 cm} & & \hspace{-0.5cm} ds^2 = \nonumber \\
     & & \hspace{-0.5cm}    
   -{\biggl [}\alpha^2 (\rho^2 + r_+^2)-2M\Pi 
   [\alpha^2 (\rho^2 + r_+^2)]^{-1/2} +4Q^2_{\rm e}
     [\alpha^4 (\rho^2 + r_+^2)]^{-1} {\biggr ]} dt^2
                                             \nonumber \\
    & &   \hspace{-0.5cm}   
         -\frac{8}{3}P{\biggl [}[\alpha^2 
        (\rho^2 + r_+^2)]^{-1/2}  -\frac{2Q^2_{\rm e}}
        {M \Pi}[\alpha^4 (\rho^2 + r_+^2)]^{-1}  
        {\biggr ]} 2dt dz  
                                                 \nonumber \\
    & &     \hspace{-0.5cm} 
       + \frac{\frac{\rho^2}{(\rho^2 + r_+^2)}}
    { {\biggl [}  \alpha^2 (\rho^2 + r_+^2) +2M (4-3\Pi)     
     [\alpha^2 (\rho^2 + r_+^2)]^{-1/2}
     -4\chi_{\rm m}^2 [\alpha^2 (\rho^2 + r_+^2)]^{-1} 
     {\biggr ]}}           d\rho^2
                                             \nonumber \\
    & &   \hspace{-0.5cm}
      +  (\rho^2 + r_+^2) d\varphi^2  \nonumber \\
    & &  \hspace{-0.5cm} 
   + {\biggl [}\alpha^2 (\rho^2 + r_+^2)      
      +\frac{4M(2-\Pi)}{[\alpha^2 (\rho^2 + r_+^2)]^{1/2}}
    -\frac{8(2-\Pi)}{4-3\Pi}\frac{\chi_{\rm m}^2}   
     {[\alpha^2 (\rho^2 + r_+^2)]}      
         {\biggr ]}   dz^2 \:,
                           \label{Met_MJQ_2}  
\end{eqnarray}
}where, the mass per unit length,  
the linear momentum per unit length and the electric charge 
per unit length of the boosted solution are
$M =b(\gamma^2+2\lambda^2 / \alpha^2)/8$,  
$P= 3b\gamma\lambda/(8\alpha)$ and
$Q_{\rm e} = \lambda   \chi_{\rm m}$.
We have defined the parameter $\Pi$ as
$\Pi \equiv 1- \sqrt{1-8P^2/(9M^2)}\,$ and
the condition that $\Pi$ remains real imposes a 
restriction on the allowed values of the linear momentum:
$P^2 \leq \frac{8}{9}M^2$.
The vector potential is given by
\begin{equation}
A=\frac{ A(\rho)}{\sqrt{8-6\Pi}}{\biggl [}
   - \sqrt{2\Pi}dt +2\sqrt{2-\Pi}  dz {\biggr ]}\:,
                                    \label{VEC_POTENT_P}
\end{equation}
where 
$A(\rho)=-[2 \chi_{\rm m}/(\alpha^3 r_+)] 
\ln[2(r_+ + \sqrt{\rho^2+r_+^2})/\rho]$.
Notice that when one  applies the 
translation boost in the $t$-$z$ plane to endow our 
initial static spacetime with linear momentum, there 
appears  a net electric charge proportional to   
$\lambda   \chi_{\rm m}$.
The static solution can be obtained by putting 
$\Pi=0$ (and thus  $P=0$ and $Q_{\rm e}=0$) on the above 
expressions.

\vskip 1cm
\noindent
{\small
\section{\bf Conclusions}
}
\vskip 3mm

We have added a Maxwell term to  Einstein Relativity with 
$\Lambda<0$. For the static spacetime, the electric and 
magnetic fields cannot be simultaneously non-zero, i.e. there 
is no static dyonic solution. 
Pure electrically charged solutions of the theory have been 
studied in detail in \cite{Zanchin_Lemos}. 

We have found two families of solutions. One yields a 
spacetime with longitudinal
magnetic field [the only non-vanishing component of the 
vector potential is $A_{\varphi}(\rho)$] 
generated by a static magnetic line source 
(an infinite straight magnetic string or a closed magnetic string). 
The corresponding spinning magnetic source that produces
in addition a radial electric field was also found.
The source for the longitudinal magnetic field solution can be 
interpreted as  composed by a system of two symmetric and 
superposed electrically charged lines with one of the 
electrically charged lines being at rest and the other 
spinning.
The other solution gives a spacetime with 
angular magnetic field [$A_z(\rho) \neq 0$].
This angular magnetic field solution can be similarly 
interpreted as composed by two electrically charged lines 
but now one is at rest and the other has a velocity along the 
axis. This solution cannot be extended down to the origin. 
The two families of solutions are asymptotically 
anti-de Sitter which allowed us to find the mass, momentum 
and electric charge of the solutions.

As we said, there is a relation between 
cylindrically symmetric four 
dimensional solutions and spacetimes generated by point 
sources in three dimensions.
Therefore, we expect that three dimensional analogue
of the solutions presented in this paper exists.
The dimensional reduction of  the longitudinal magnetic 
field solution yields a Brans-Dicke theory 
(see \cite{OscarLemos_2}). 
On the other hand 
the dimensional reduction of  the angular magnetic 
field solution yields an effective three dimensional
theory which is not a pure Brans-Dicke theory
since it has an extra gauge field 
(see \cite{Zanchin_Lemos_2}).

\vskip .5cm
\section*{Acknowledgments} 
This work was partially funded by Funda\c c\~ao para a 
Ci\^encia e Tecnologia (FCT) through 
project PESO/PRO/2000/4014. 
OJCD also acknowledges finantial support from the portuguese FCT 
through PRAXIS XXI programme. JPSL thanks Observat\'orio 
Nacional do Rio de Janeiro for hospitality.

\vskip 1cm

\end{document}